\documentclass[twocolumn,prb]{revtex4}
\usepackage{graphicx}
\usepackage{dcolumn}
\usepackage{bm}
\usepackage{amsmath}

\begin{document}

\preprint{}
\title{Nonmonotonic temperature dependence of critical current in diffusive $
d$-wave junctions}
\author{T. Yokoyama$^1$, Y. Tanaka$^1$, A. A. Golubov$^2$ and Y. Asano$^3$}
\affiliation{$^1$Department of Applied Physics, Nagoya University, Nagoya, 464-8603, Japan%
\\
and CREST, Japan Science and Technology Corporation (JST) Nagoya, 464-8603,
Japan \\
$^2$ Faculty of Science and Technology, University of Twente, 7500 AE,
Enschede, The Netherlands\\
$^3$Department of Applied Physics, Hokkaido University,Sapporo, 060-8628,
Japan}
\date{\today}

\begin{abstract}
We study the Josephson effect in D/I/DN/I/D junctions, 
where I, DN and D denote an insulator, a diffusive
normal metal and a $d$-wave superconductor, respectively.
The
Josephson current is calculated based on the quasiclassical Green's function theory
with a general boundary condition for unconventional
superconducting junctions.
In contrast to $s$-wave junctions, the
product of the Josephson current and the normal state resistance is
enhanced by making the interface barriers stronger. The Josephson current has
a nonmonotonic temperature dependence due to the competition between
the proximity effect and the midgap Andreev resonant states.
\end{abstract}

\maketitle




%

%




Since the discovery of Josephson effect~\cite{Josephson} in superconductor /
insulator / superconductor (SIS) junctions, it has been studied in various
types of structures~\cite{Likharev,Golubov}. The Josephson current flows not
only through thin insulators but also through  normal metals or ferromagnets~\cite{Josephson,Likharev,Golubov,Ryazanov}.
In superconductor / diffusive normal metal / superconductor (S/DN/S)
junctions, Josephson current is carried by Cooper pairs penetrating into the
DN as a result of the proximity effect. In $s$-wave superconductor junctions, the maximum amplitude of Josephson current ($I_{C}$) monotonically increases
with the decrease of temperature~\cite{Ambegaokar,Kupriyanov,Zaitsev}. This
is a natural consequence of the fact that interference effects are stronger
at lower temperatures. The Josephson effect, however, depends strongly on
pairing symmetries of superconductors because it is an essentially phase
sensitive phenomenon. In $d$-wave superconductor / insulator / $d$-wave
superconductor (DID) junctions, for instance, $I_{C}$ first increases with
the decrease of temperature then decreases and can 
even change its sign at low temperatures for certain orientations~\cite{Barash,Golubov2,Ilichev,Testa}. 
This nonmonotonic behavior of Josephson current is caused by 0-$\pi $
transition due to the  mid-gap Andreev resonant states (MARS) forming at
junction interfaces~\cite{Buch,Tanaka95}. Similar effect is also observed in
Josephson current through ferromagnets~\cite{Ryazanov}. Thus the
nonmonotonic temperature dependence of $I_{C}$ is a sign of unusual
interference effect.

The quasiclassical Green's function theory is a powerful tool to study
quantum transport in superconducting junctions. The circuit theory~\cite%
{Nazarov2} provides the boundary conditions for the Usadel equations~\cite%
{Usadel} widely used in diffusive superconducting junctions. These boundary
conditions generalize the Kupriyanov-Lukichev conditions~\cite{Kupriyanov} 
for an arbitrary type of connector which couples the diffusive nodes.
Recently Tanaka \textit{et al.}~\cite{Nazarov2003,TNGK,p-wave} have
extended the circuit theory~\cite{Nazarov2} to systems with unconventional
superconductors. An application of the extended circuit theory to DN/$d$%
-wave superconductor junctions has revealed a strong competition
(destructive interference) between the MARS and the proximity effect in DN~%
\cite{Nazarov2003,TNGK}. This competition, however, has not yet tested
experimentally. Thus it's important to propose the way to verify this
prediction.

In the present paper, we show that Josephson effect is a suitable
tool to observe the above competition. We extend the previous theory
\cite{Nazarov2003,TNGK} in order to treat the external phase
difference between the left and right
superconductors. Applying this formalism, we calculate Josephson current in $%
d$-wave superconductor /insulator/ diffusive normal metal /insulator/ $d$-wave
superconductor (D/I/DN/I/D) junctions, solving the Usadel equations with
new boundary conditions. This allows us to study the
influences of the proximity effect and the formation of MARS on the
Josephson current simultaneously. We find that the competition between the
proximity effect and the formation of MARS provides a new mechanism
for a nonmonotonic temperature dependence of the maximum Josephson
current.

\begin{figure}[htb]
\begin{center}
\scalebox{0.4}{
\includegraphics[width=21.0cm,clip]{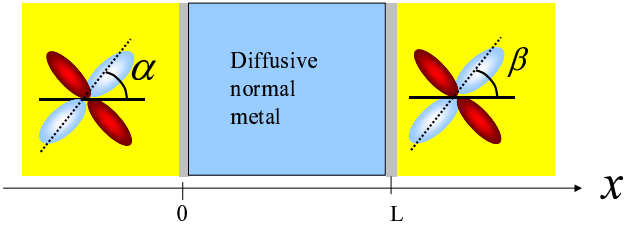}}
\end{center}
\caption{ (color online) Schematic illustration of the model.}
\label{f0}
\end{figure}

We consider a junction consisting of $d$-wave superconducting reservoirs (D)
connected by a quasi-one-dimensional diffusive conductor (DN) with a
resistance $R_{d}$ and a length $L$ much larger than the mean free path. The
DN/D interface located at $x=0$ has the resistance $R_{b}^{\prime }$, while
the DN/D interface located at $x=L$ has the resistance $R_{b}$. We model
infinitely narrow insulating barriers by the delta function $U(x)=H\delta
(x-L)+H^{\prime }\delta (x)$. The resulting transparencies of the junctions $%
T_{m}$ and $T_{m}^{\prime }$ are given by $T_{m}=4\cos ^{2}\phi /(4\cos
^{2}\phi +Z^{2})$ and $T_{m}^{\prime }=4\cos ^{2}\phi /(4\cos ^{2}\phi +{%
Z^{\prime }}^{2})$, where $Z=2H/v_{F}$ and $Z^{\prime }=2H^{\prime }/v_{F}$
are dimensionless constants and $\phi $ is the injection angle measured from
the interface normal to the junction and $v_{F}$ is Fermi velocity. We show
the schematic illustration of the model in Fig. \ref{f0}.

We parameterize the quasiclassical Green's functions $G$ and $F$ with a
function $\Phi _{\omega }$ \cite{Likharev,Golubov}:
\begin{equation}
G_{\omega }=\frac{\omega }{{\sqrt{\omega ^{2}+\Phi _{\omega }\Phi _{-\omega
}^{\ast }}}},F_{\omega }=\frac{{\Phi _{\omega }}}{{\sqrt{\omega ^{2}+\Phi
_{\omega }\Phi _{-\omega }^{\ast }}}}
\end{equation}%
where  $\omega $ is the Matsubara frequency. Then Usadel equation reads\cite%
{Usadel}
\begin{equation}
\xi ^{2}\frac{{\pi T_{C}}}{{\omega G_{\omega }}}\frac{\partial }{{\partial x}%
}\left( {G_{\omega }^{2}\frac{\partial }{{\partial x}}\Phi _{\omega }}%
\right) -\Phi _{\omega }=0
\end{equation}%
with the coherence length $\xi =\sqrt{D/2\pi T_{C}}$, the diffusion constant
$D$ and the transition temperature $T_{C}$. From the conservation law for the 
matrix current\cite{TNGK}, the boundary conditions are given by 
\begin{equation*}
\frac{{G_{\omega }}}{\omega }\frac{\partial }{{\partial x}}\Phi _{\omega }=-%
\frac{{R_{d}}}{{R_{b}^{\prime }L}}\left( {-\frac{{\Phi _{\omega }}}{\omega }%
I_{1}^{\prime }+e^{-i\varphi }\left( {I_{2}^{\prime }-iI_{3}^{\prime }}%
\right) }\right)
\end{equation*}%
\begin{equation*}
I_{1}^{\prime }=\left\langle {\frac{{2T_{m}^{\prime }g_{S}^{\prime }}}{{%
A^{\prime }}}}\right\rangle ^{\prime }I_{2}^{\prime }=\left\langle {\frac{{%
2T_{m}^{\prime }\overline{f}_{S}^{\prime }}}{{A^{\prime }}}}\right\rangle
^{\prime }I_{3}^{\prime }=\left\langle {\frac{{2T_{m}^{\prime }f_{S}^{\prime
}}}{{A^{\prime }}}}\right\rangle ^{\prime }
\end{equation*}%
\begin{equation*}
A^{\prime }=2-T_{m}^{\prime }+T_{m}^{\prime }(g_{S}^{\prime }G_{\omega }
\end{equation*}%
\begin{equation*}
+\overline{f}_{S}^{\prime }(B^{\prime }\cos \varphi +C^{\prime }\sin \varphi
)+f_{S}^{\prime }(C^{\prime }\cos \varphi -B^{\prime }\sin \varphi ))
\end{equation*}%
\begin{equation*}
B^{\prime }=\frac{{G_{\omega }}}{{2\omega }}\left( {\Phi _{\omega }+\Phi
_{-\omega }^{\ast }}\right) \quad C^{\prime }=\frac{{iG_{\omega }}}{{2\omega
}}\left( {\Phi _{\omega }-\Phi _{-\omega }^{\ast }}\right)
\end{equation*}%
\begin{equation*}
g_{S}^{\prime }=\frac{{g_{+}+g_{-}}}{{1+g_{+}g_{-}+f_{+}f_{-}}}\quad
\overline{f}_{S}^{\prime }=\frac{{f_{+}+f_{-}}}{{1+g_{+}g_{-}+f_{+}f_{-}}}
\end{equation*}%
\begin{equation}
f_{S}^{\prime }=\frac{{i\left( {f_{+}g_{-}-f_{-}g_{+}}\right) }}{{%
1+g_{+}g_{-}+f_{+}f_{-}}} \label{bc}
\end{equation}%
$g_{\pm }=\omega /\sqrt{\omega ^{2}+\Delta _{_{\pm }}^{2}}$ $f_{\pm }=\Delta
_{\pm }/\sqrt{\omega ^{2}+\Delta _{_{\pm }}^{2}}$ with $\Delta _{\pm }=\Delta
(T)\cos (2\phi \mp 2\alpha )$ at $x=0$ and
\begin{equation}
\frac{{G_{\omega }}}{\omega }\frac{\partial }{{\partial x}}\Phi _{\omega }=%
\frac{{R_{d}}}{{R_{b}L}}\left( {-\frac{{\Phi _{\omega }}}{\omega }%
I_{1}+\left( {I_{2}-iI_{3}}\right) }\right)
\end{equation}%
at $x=L$, where $I_{1},I_{2}$ and $I_{3}$ are defined similarly to $%
I_{1}^{\prime },$ $I_{2}^{\prime }$ and $I_{3}^{\prime }$ by removing the
superscript '$^{\prime }$', exchanging subscript'+' for subscript'-', putting $\varphi =0$ and
substituting $\beta $ for $\alpha $, respectively. Here $\varphi $ is the
external phase difference across the junctions, and $\alpha $ and $\beta $
denote the angles between the normal to the interface and the crystal axes
of $d$-wave superconductors for $x\leq 0$ and $x\geq L$ respectively.

These boundary conditions derived above are quite general since with a proper
choice of $\Delta _{\pm }$ they are applicable to any unconventional
superconductor with $S_{z}=0$ in a time reversal symmetry conserving state.
Here, $S_{z}$ denotes the $z$-component of the total spin of a Cooper pair.

The average over the various angles of injected particles at the interface
is defined as
\begin{equation}
<B(\phi)>^{(\prime)} = \frac{\int_{-\pi/2}^{\pi/2} d\phi \cos\phi B(\phi)}{
\int_{-\pi/2}^{\pi/2} d\phi T^{(\prime)}(\phi)\cos\phi}  \label{aave}
\end{equation}
with $B(\phi)=B$ and $T^{(\prime)}(\phi)=T_{m}^{(\prime)}$.  It is important
to note that the solution of the Usadel equation is invariant under the
transformation $\alpha \to -\alpha$ or $\beta \to -\beta$. This is clear by
replacing $\phi$ with $-\phi$ in the angular averaging. The resistance of
the interface $R_{b}^{(\prime)}$ is given by
\begin{equation}
R_{b}^{(\prime)}=R_{0}^{(\prime)} \frac{2} {\int_{-\pi/2}^{\pi/2} d\phi
T^{(\prime)}(\phi)\cos\phi}.
\end{equation}
Here, for example, $R_{b}^{(\prime)}$ denotes $R_{b}$ or $R_{b}^{\prime}$,
and $R_{0}^{(\prime)}$ is Sharvin resistance, which in three-dimensional
case is given by $R_{0}^{(\prime)-1}=e^{2}k_{F}^2S_c^{(\prime)}/(4\pi^{2} )$%
, where $k_{F}$ is the Fermi wave-vector and $S_c^{(\prime)}$ is the
constriction area.

The Josephson current is given by 
\begin{equation}
\frac{{eIR}}{{\pi T_C }} = i\frac{{RTL}}{{2R_d T_C }}\sum\limits_\omega {%
\frac{{G_\omega ^2 }}{{\omega ^2 }}} \left( {\Phi _\omega \frac{\partial }{{%
\partial x}}\Phi _{ - \omega }^ * - \Phi _{ - \omega }^ * \frac{\partial }{{%
\partial x}}\Phi _\omega } \right)
\end{equation}
with temperature $T$ and $R \equiv R_{d}+R_{b}+R_{b}^{\prime}$. In the following we focus on the $I_C R$ value where $I_C$
denotes the magnitude of the maximum Josephson current. We consider symmetric junctions with $R_{b}=R^{\prime }_{b}$ and
$Z=Z^{\prime }$ for simplicity. 



\begin{figure}[tbh]
\begin{center}
\scalebox{0.4}{
\includegraphics[width=22.0cm,clip]{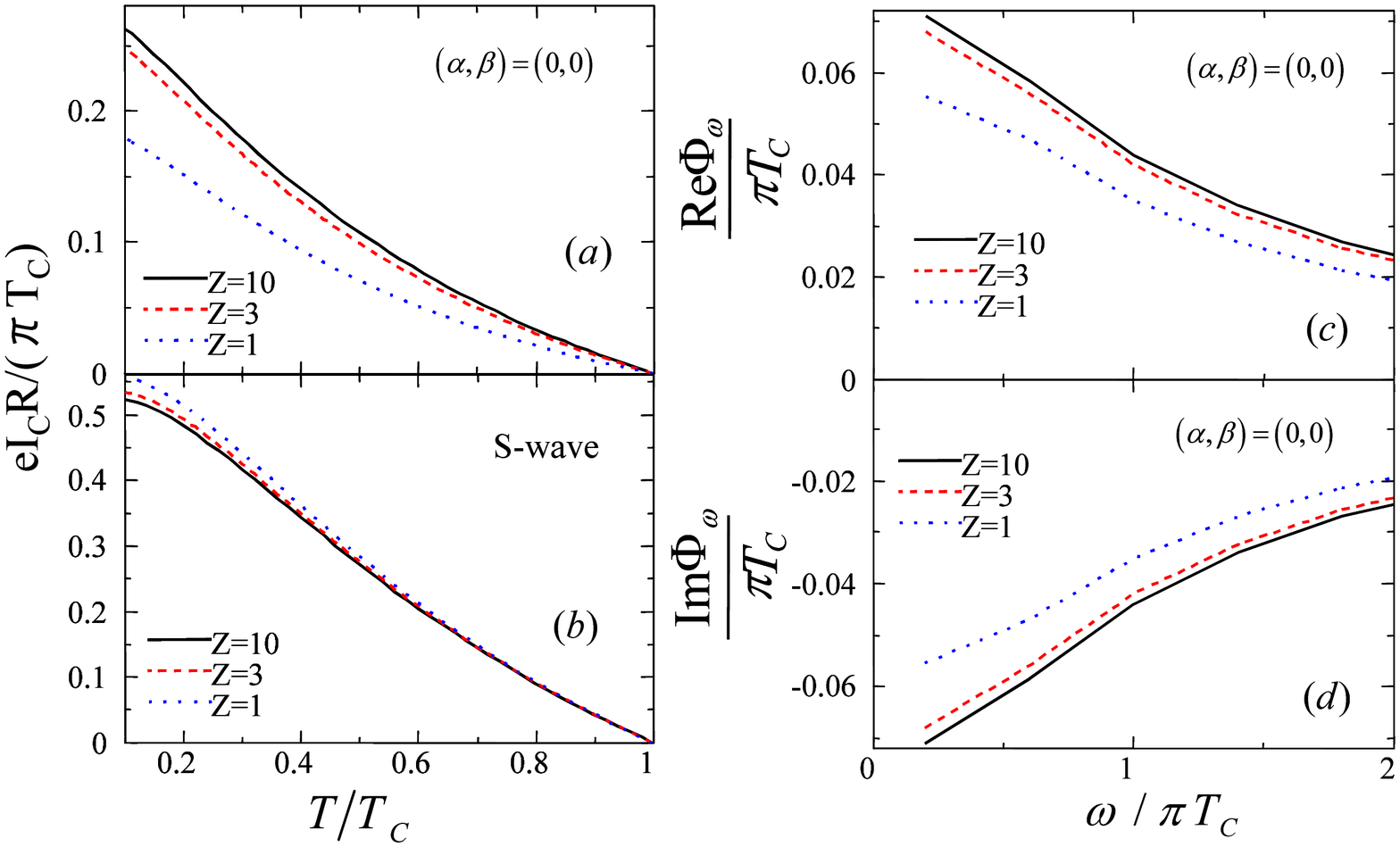}
}
\end{center}
\caption{(color online) (a) Temperature dependence of $I_{C}R$ for $\left( {%
\protect\alpha ,\protect\beta }\right) =\left( {0,0}\right) $. (b) The
results for $s$-wave junctions for comparison. Real (c) and imaginary
(d) parts of $\Phi _{\protect\omega }$ as a function of $\protect\omega$. Here we choose $R_{d}/R_{b}=1$ and $E_{Th}/\Delta (0)=1$. }
\label{f2}
\end{figure}

In Fig.~\ref{f2}(a), $I_{C}R$ is plotted as a function of temperature for $%
R_{d}/R_{b}=1$, $E_{Th}/\Delta (0)=1$, and $\left( {\alpha ,\beta }\right)
=\left( {0,0}\right) $. The results show that $I_{C}R$ increases with the
decrease of $T$ as it does in $s$-wave junctions as shown in (b). Amplitude of $%
I_{C}R$ increases with the increase of $Z$ in $d$-wave junctions, whereas
the opposite tendency can be  seen in $s$-wave junctions. The sign change of pair
potential is responsible for the $Z$ dependence of $I_CR$ in (a). In the $d$%
-wave symmetry with $\alpha=\beta=0$, injection angles of a quasiparticle
can be divided into two regions: $\phi_+= \{ \phi | 0 \leq |\phi|< \pi/4 \}$
and $\phi_- =\{ \phi | \pi/4 \leq |\phi| \leq \pi/2 \} $. The sign of pair
potential for $\phi_+$ and that for $\phi_-$ are opposite to each other.
 In general, such a sign change of pair potentials leads to the
suppression of the proximity effect in DN and hence Josephson currents. This is
because Cooper pairs originated from the positive part of pair potentials
cancel those originated from the negative part of pair potentials.
For small $Z$, a quasiparticle with $\phi_+$ and that with $\phi_-$ give
opposite contribution to the proximity effect. In the actual calculation,
the cancellation due to the sign change of the pair potential reduces the
magnitudes of $I_{2}$, $I_3$, $I_{2}^\prime$, and $I_3^\prime$. On the other hand for large $Z$,
integration over $\phi_+$ has the dominant contribution to $I_{2}$, $I_3$, $%
I_{2}^\prime$, and $I_3^\prime$. As a result, the cancellation of Cooper
pairs becomes negligible in low transparent junctions. Thus $I_CR$ in the $d$%
-wave junctions increases with increasing $Z$. This argument can be
confirmed by the behavior of $\Phi _{\omega }$ which represents a degree of the
proximity effect in DN. In Figs.~\ref{f2}(c) and (d), we show $\Phi _{\omega
}$ as a function of $\omega$, where we choose $x=L/2$, $\varphi =\pi /2$, $%
T/T_{C}=0.2$, and $(\alpha,\beta)=(0,0)$. The magnitudes of  $Re\, \Phi _{\omega }$ and $%
Im \, \Phi _{\omega }$ increase with increasing $Z$.
This is the reason of the enhancement of $I_{C}R$ product for low
transparent junctions with large magnitude of $Z$.


\begin{figure}[tbh]
\begin{center}
\scalebox{0.4}{
\includegraphics[width=22.0cm,clip]{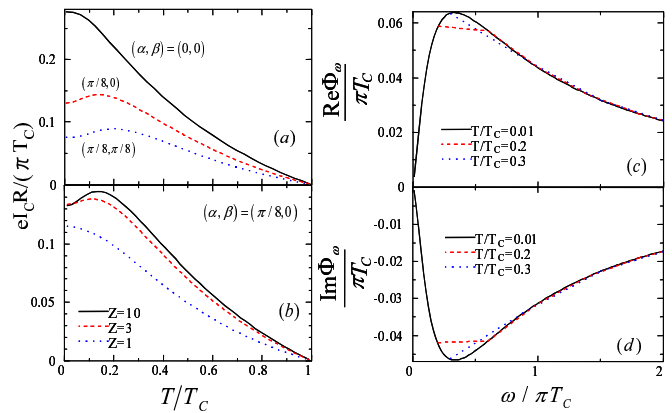}
}
\end{center}
\caption{(color online) Temperature dependence of (a) $I_{C}R$ at $Z=10$ for
several orientation angles. (b) $I_{C}R$ at $\left( {\protect\alpha ,\protect%
\beta }\right) =\left( {\protect\pi /8,0}\right) $ for several $Z$ values.
Real and imaginary parts of $\Phi _{\protect\omega }$ as a function of $\protect\omega$ for $\left( {\protect%
\alpha ,\protect\beta }\right) =\left( {\protect\pi /8,0}\right) $ and $\protect\varphi =\protect\pi /2$ at $x=L/2$ are shown 
 in (c) and (d) respectively. Here we choose $R_{d}/R_{b}=1$ and $E_{Th}/\Delta (0)=1$. }
\label{f7}
\end{figure}

Next we focus on the Josephson effect with nonzero values of $\alpha $ and $%
\beta $. Figure~\ref{f7} (a) displays $I_{C}R$ as a function of temperature
for $Z=10$, $R_{d}/R_{b}=1$ and $E_{Th}/\Delta (0)=1$. In contrast to $I_{C}R
$ for $\left( {\alpha ,\beta }\right) =\left( {0,0}\right) $, the results
for $\left( {\alpha ,\beta }\right) =\left( {\pi /8,0}\right) $ and $\left( {%
\pi /8,\pi /8}\right) $ show a nonmonotonic temperature dependence. The
transparency at the interfaces greatly changes the peak structure as shown in
(b), where temperature dependence of $I_{C}R$ is plotted for several $Z$ at $%
\left( {\alpha ,\beta }\right) =\left( {\pi /8,0}\right) $. The peak
structure is smeared as the decrease of $Z$.

To understand the origin of the nonmonotonic temperature dependence of $%
I_{C}R$, we have to consider in detail the mechanism of competition between
the proximity effect and the formation of the MARS. For example, at $\alpha =\beta =\pi
/8$, the injection angles of a quasiparticle are separated into four
regions: (i) $\phi _{M}=\{\phi |\pi /8\leq \phi <3\pi /8\}$, (ii) $\phi
_{P}=\{\phi |0\leq \phi <\pi /8$ and $3\pi /8\leq \phi \leq \pi /2\}$, (iii)
$-\phi _{M}$, and (iv) $-\phi _{P}$. Transport channels within $\pm \phi _{P}
$ contribute to the proximity effect but do not form
the MARS. On the other hand, transport channels within $\pm \phi _{M}$ are
responsible for the formation of the MARS but do not contribute to the
proximity effect. Thus channels within $\pm \phi _{P}$ and those within $\pm
\phi _{M}$ play different roles in electron transport. At the same time, the
types of wave transmission are different for these two transport channels.
The channels within $\pm \phi _{P}$ are open for the normal transmission. On
the other hand, in the channels within $\pm \phi _{M}$, the resonant
transmission is also possible for small $\omega $. At high temperatures,
effects of the MARS are negligible and Josephson currents are carried
through channels within $\pm \phi _{P}$. At low temperatures, the resonant
transmission through $\pm \phi _{M}$ governs electric currents. This is a
result of an important property of resonant transport. When a normal
transport channel and a resonant transport channel are available in
parallel, a quasiparticle tends to choose the resonant channel for its
transmission at low temperatures. This property can be confirmed in the present junctions by $
\omega $ dependence of $\Phi $ as shown in Fig.~\ref{f7} (c) and (d). The results 
show that $\Phi $ rapidly decreases to zero in the limit of $\omega =0$,
which implies no diffusion of Cooper pairs into DN through the normal
channels within $\pm \phi _{P}$. At $\omega =0$, quasiparticles choose the
resonant channels within $\pm \phi _{M}$. Electric currents through $\pm \phi
_{M}$, however, do not contribute to the net Josephson current because of
cancellation of the currents through $\phi _{M}$ by those through $-\phi _{M}
$. Thus $\Phi $ goes to zero at $\omega =0$, which leads to the suppression
of the Josephson current at low temperatures and hence the nonmonotonic
temperature dependence of $I_{C}$. The influence of the MARS is more
remarkable for larger $Z$. Therefore the resulting nonmonotonic temperature
dependence becomes much more pronounced for large $Z$ as shown in Fig.~\ref%
{f7}(b). The origin of the nonmonotonic temperature dependence of $%
I_{C}R$ in D/I/DN/I/D junctions is different from that in the mirror-type DID
junctions (i.e., $\alpha =-\beta $), where the current through the resonant
channels causes the 0-$\pi $ transition and the nonmonotonic temperature
dependence~\cite{Barash,Ilichev,Testa}.

It is important to note that the angle average in Eq.~(\ref{aave})
at the two interfaces can be carried out independently. The physical meaning
can be explained by considering the motion of a quasiparticle which starts
from the left interface with a certain injection angle and reaches the right
interface after traveling across the  DN. At the right interface, injection
angles of a quasiparticle are independent of the initial injection angle at
the left interface because of the diffusive scatterings in the DN. This property
allows one to carry out the angular averaging at the two interfaces
independently. When a DN is replaced by a clean normal metal or a clean
insulator, the injection angles at both interfaces are correlated to each
other. This is because the injection angle of a quasiparticle at the left
interface is conserved when the quasiparticle reaches the right interface.
As a result, in the clean limit, Josephson current in a symmetric
junction (i.e., $\alpha =\beta $) has a behavior different from that in a
mirror-type junction with $\alpha =-\beta $. In a symmetric DID junction, $%
I_{C}R$ increases monotonically with decreasing temperature~\cite%
{Barash,Golubov2}. In the D/I/DN/I/D junctions considered in the present paper,
the mirror-type and the symmetric configurations yield the same Josephson
current because of the independent angle averaging at the two interfaces.

\begin{figure}[tbh]
\begin{center}
\scalebox{0.4}{
\includegraphics[width=22.0cm,clip]{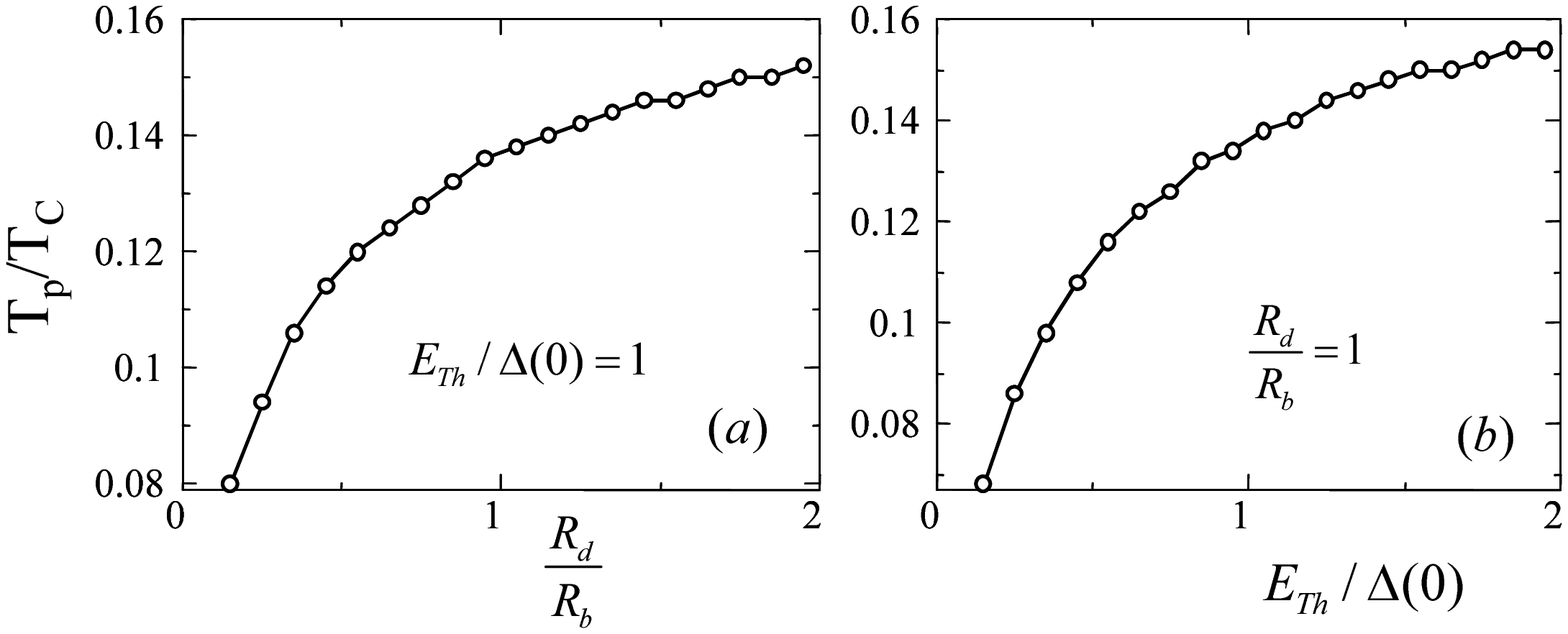}
}
\end{center}
\caption{The peak temperature as a function of $R_{d}/R_{b}$ (a) and $%
E_{Th}/\Delta (0)$ (b) for $Z=10$ and $\left( {\protect\alpha ,\protect\beta
}\right) =\left( {\protect\pi /8,0}\right) $.}
\label{f8}
\end{figure}

To study the nonmonotonic temperature dependence of $I_{C}R$ further, we
define the temperature $T_{p}$ at which $I_{C}R$ has a maximum. We have
confirmed that $T_{p}$ is insensitive to $Z$ when $Z$ is sufficiently large.
The calculated $T_{p}$ shown in Fig.~\ref{f8} increases monotonically with
the increase of $R_{d}/R_{b}$ and $E_{Th}/\Delta (0)$. These results imply
that large magnitudes of $Z$, $R_{d}/R_{b}$ and $E_{Th}/\Delta (0)$ are
needed to observe the peak structure experimentally. These conditions are
satisfied in junctions with high insulating barriers, small cross section
area of the DN and thin DN.

In the present paper, we have studied the Josephson effect in $d$-wave
superconductor / insulator / diffusive normal metal / insulator /$d$ -wave
superconductor junctions by solving the Usadel equations. We have derived
the boundary conditions for the quasiclassical Green's function which make it possible to
calculate the Josephson current across the structure. By calculating the
Josephson current for various parameters, we have clarified the following
points. (1) In contrast to the conventional $s$-wave junctions, the
magnitude of $I_{C}R$ is enhanced with the increase of the insulating
barrier strengths at the interfaces. (2) The $I_{C}R$ value may exhibit a
nonmonotonic temperature dependence due to the competition between the
proximity effect and the formation of the MARS. The origin of such a
nonmonotonic behavior is completely different from that in the established
case of a clean DID junction~\cite{Barash,Ilichev, Testa} where $0$-$\pi $
transition occurs due to the MARS. In order to observe experimentally the
predicted nonmonotonic temperature dependence of $I_{C}$ in D/I/DN/I/D
junctions, large magnitudes of  parameters $Z$, $R_{d}/R_{b}$,
and $E_{Th}/\Delta (0)$ are required. Note that we can reproduce the main
conclusions obtained here within the quasiclassical theory of
superconductivity by numerical simulations based on the
recursive Green function method\cite{Asano}.

It follows from the present approach that similar nonmonotonic temperature
dependence is also expected for $s$-wave superconductor  / insulator / diffusive normal metal / insulator / $d$ -wave superconductor junctions. To the best of our knowledge of Josephson junctions consisting of $s$- and $d$ -wave superconductors, 
there is only one report in literature about the observation of nonmonotonic
temperature dependence of $I_{C}$ in YBaCuO/Pb junctions~\cite%
{Iguchi}. Though this experiment might be possibly relevant to our results,
more experimental data are needed to check whether it is the case.

%

%



\begin{thebibliography}{99}
\bibitem{Josephson} B. D. Josephson, Phys. Lett. \textbf{1}, 251 (1962).

\bibitem{Likharev} K.K. Likharev, Rev. Mod. Phys. \textbf{51}, 101 (1979).

\bibitem{Golubov} A. A. Golubov, M. Yu. Kupriyanov, and E. Il$^{\prime}$%
ichev Rev. Mod. Phys. \textbf{76}, 411 (2004).

\bibitem{Ryazanov} V. V. Ryazanov, V. A. Oboznov, A. Yu. Rusanov, A. V.
Veretennikov, A. A. Golubov, and J. Aarts, Phys Rev. Lett. \textbf{86}, 2427
(2001).

\bibitem{Ambegaokar} V. Ambegaokar and A. Baratoff, Phys. Rev. Lett. \textbf{%
10}, 486 (1963).

\bibitem{Kupriyanov} M. Yu. Kupriyanov and V. F. Lukichev, Zh. Exp. Teor.
Fiz. \textbf{94} (1988) 139 [Sov. Phys. JETP \textbf{67}, (1988) 1163].

\bibitem{Zaitsev} A. V. Zaitsev, Physica C \textbf{185-189}, 2539 (1991).

\bibitem{Barash} Yu. S. Barash, H. Burkhardt, and D. Rainer, Phys. Rev.
Lett. \textbf{77}, 4070 (1996); Y. Tanaka and S. Kashiwaya, Phys. Rev. B
\textbf{53}, R11957 (1996); Y. Tanaka and S. Kashiwaya, Phys. Rev. B \textbf{%
56}, 892 (1997).

\bibitem{Golubov2} A. A. Golubov and M. Yu. Kupriyanov, JETP Lett. \textbf{69%
}, 262 (1999).

\bibitem{Ilichev} E. Il'ichev, M. Grajcar, R. Hlubina, \emph{et al.}, Phys. Rev. Lett. \textbf{%
86}, 5369 (2001).

\bibitem{Testa} G. Testa, E. Sarnelli, A. Monaco, E. Esposito, M. Ejrnaes, D.-J. Kang, S. H. Mennema, E. J. Tarte, and M. G. Blamire
Phys. Rev. B \textbf{71}, 134520 (2005)


\bibitem{Buch} L.J. Buchholtz and G. Zwicknagl, Phys. Rev. B \textbf{23}
5788 (1981); J. Hara and K. Nagai, Prog. Theor. Phys. \textbf{74} (1986)
1237 ; C. Bruder, Phys. Rev. B \textbf{41}, (1990) 4017; C.R. Hu, Phys. Rev.
Lett. \textbf{72}, (1994) 1526.

\bibitem{Tanaka95} Y. Tanaka and S. Kashiwaya, Phys. Rev. Lett. \textbf{74},
(1995) 3451.

\bibitem{Nazarov2} Yu. V. Nazarov, Superlattices and Microstructuctures
\textbf{25}, 1221 (1999).

\bibitem{Usadel} K.D. Usadel, Phys. Rev. Lett. \textbf{25}, 507 (1970).

\bibitem{Nazarov2003} Y. Tanaka, Y.V. Nazarov and S. Kashiwaya, Phys. Rev.
Lett. \textbf{90}, 167003 (2003).

\bibitem{TNGK} Y. Tanaka, Yu. V. Nazarov, A. A. Golubov, and S. Kashiwaya,
Phys. Rev. B \textbf{69}, 144519 (2004).

\bibitem{p-wave} Y. Tanaka and S. Kashiwaya, Phys. Rev. B \textbf{70},
012507 (2004); Y. Tanaka, S. Kashiwaya and T. Yokoyama, Phys. Rev. B \textbf{%
\ 71}, 094513 (2005); Y. Tanaka, Y. Asano, A. A. Golubov and S. Kashiwaya,
Phys. Rev. B \textbf{72}, 140503(R) (2005).


\bibitem{Asano} Y. Asano, Phys. Rev. B \textbf{64}, 014511 (2001); J. Phys.
Soc. Jpn \textbf{71}, 905 (2002).


\bibitem{Iguchi} I. Iguchi and Z. Wen, Phys. Rev. B \textbf{49}, R12388
(1994).
\end{thebibliography}
\end{document}